\documentclass[12pt,a4paper,epsf]{article}
\usepackage{graphics,amssymb,amsmath,xcolor,cite,longtable,slashed}
\usepackage[dvips]{lscape,graphicx}

\usepackage[pagewise, displaymath, mathlines]{lineno}

\usepackage{draftwatermark}
\SetWatermarkText{}
\SetWatermarkScale{6}
\SetWatermarkLightness{0.9}

\newcommand{\ct}{\cite}
\newcommand{\lb}{\label}

\newcommand{\bc}{\begin{center}}
\newcommand{\ec}{\end{center}}
\newcommand{\bd}{\begin{displaymath}}
\newcommand{\ed}{\end{displaymath}}
\newcommand{\be}{\begin{equation}}
\newcommand{\ee}{\end{equation}}
\newcommand{\ba}{\begin{array}}
\newcommand{\ea}{\end{array}}
\newcommand{\bea}{\begin{eqnarray}}
\newcommand{\eea}{\end{eqnarray}}
\newcommand{\bt}{\begin{tabular}}
\newcommand{\et}{\end{tabular}}
\newcommand{\un}{\underline}

\newcommand{\bp}{\begin{picture}}
\newcommand{\ep}{\end{picture}}
\newcommand{\bfi}{\begin{figure}}
\newcommand{\efi}{\end{figure}}

\newcounter{firstbib}

\def\fun#1#2{\lower3.6pt\vbox{\baselineskip0pt\lineskip.9pt
\ialign{$\mathsurround=0pt#1\hfil##\hfil$\crcr#2\crcr\sim\crcr}}}

\begin{document}

\title{\LARGE \bf {Topological Structure of the Vacuum, Cosmological Constant and Dark Energy}}
\author{\large\bf B.G. Sidharth ${}^{1}$\footnote{iiamisbgs@yahoo.co.in, birlasc@gmail.com} ,
\; A. Das ${}^{1}$\footnote{parbihtih3@gmail.com} ,
\; C.R. Das ${}^{2}$\footnote{das@theor.jinr.ru} ,\\
\bf
\; L.V. Laperashvili ${}^{3}$\footnote{laper@itep.ru}
\; and H.B. Nielsen ${}^{4}$\footnote{hbech@nbi.dk}\\\\
{\large \it ${}^{1}$ International Institute of Applicable Mathematics}\\
{\large \it and Information Sciences,}\\
{\large \it B.M. Birla Science Centre}\\
{\large \it Adarsh Nagar, 500063 Hyderabad, India}\\\\
{\large \it ${}^{2}$ Bogoliubov Laboratory of Theoretical Physics}\\
{\large \it Joint Institute for Nuclear Research}\\
{\large \it International Intergovernmental Organization,}\\
{\large \it Joliot-Curie 6, 141980 Dubna, Moscow region, Russia}\\\\
{\large \it ${}^{3}$ The Institute of Theoretical and Experimental Physics,}\\
{\large\it National Research Center ``Kurchatov Institute'',}\\
{\large\it Bolshaya Cheremushkinskaya 25, 117218 Moscow, Russia}\\\\
{\large \it ${}^{4}$ Niels Bohr Institute,}\\
{\large \it Blegdamsvej, 17-21, DK 2100 Copenhagen, Denmark}}

\date{}
\maketitle

\thispagestyle{empty}

\vspace{1cm}



\thispagestyle{empty}

\newpage

\begin{abstract}

In this review we present a theory of cosmological constant and
Dark Energy (DE), based on the topological structure of the
vacuum. The Multiple Point Principle (MPP) is reviewed. It
demonstrates the existence of the two vacua into the SM. The
Froggatt-Nielsen's prediction of the top-quark and Higgs masses is
given in the assumption that there exist two degenerate vacua in
the SM. This prediction was improved by the next order
calculations. We also considered B.G. Sidharth's theory of
cosmological constant based on the non-commutative geometry of the
Planck scale space-time, what gives an extremely small DE density
providing the accelerating expansion of the Universe. Theory of
two degenerate vacua -- the Planck scale phase and Electroweak
(EW) phase -- also is reviewed, topological defects in these vacua
are investigated, also the Compton wavelength phase suggested by
B.G. Sidharth was discussed. A general theory of the phase
transition and the problem of the vacuum stability in the SM is
reviewed. Assuming that the recently discovered at the LHC new
resonance with mass $m_S \simeq 750$ GeV is a new scalar $S$ bound
state $6t + 6\bar t$, earlier predicted by C.D. Froggatt, H.B.
Nielsen and L.V. Laperashvili, we try to provide the vacuum
stability in the SM and exact accuracy of the MPP.

\end{abstract}

\newpage

\section*{Contents}

{\bf\large
\begin{enumerate}
\item[1.] Introduction.

\item[2.] Multiple Point Principle and the prediction of the top and
Higgs masses.

\item[3.] Cosmological constant and topological structure of the\\ vacuum.
\begin{enumerate}
\item[a.] Sidharth's theory of cosmological constant and Dark\\ Energy.

\item[b.] What is the  Universe vacuum?
\end{enumerate}
\item[4.] Phase transition(s) in the Universe.

\item[5.] Vacuum stability and the Multiple Point Principle.
\begin{enumerate}
\item[a.] Two-loop corrections to the Higgs mass.

\item[b.] Could the Multiple Point Principle be exact due to\\
corrections from the new bound state 6t + 6anti-t?

\item[c.] The effect from the new bound states $6t + 6\bar t$ on the\\
measured Higgs mass. The main diagrams correcting the effective
Higgs self-interaction coupling constant $\lambda$.
\end{enumerate}
\item[6.] Summary and Conclusions.
\end{enumerate}
}

\newpage

\section{Introduction}

The Standard Model (SM) is a theory with a group of symmetry:
\be  G_{SM} = SU(3)_c\times SU(2)_L\times U(1)_Y, \lb{1} \ee
which contains quarks (u, d, s, c, b, t), leptons (e, $\nu$), the
Higgs boson $H$ and gauge fields: gluons $G_{\mu}$, vector bosons
$W_{\mu}$ and $Z_{\mu}$, and electromagnetic field $A_{\mu}$. The
vast majority of the available experimental data is consistent
with the Standard Model predictions. All accelerator physics seems
to fit well with the SM, except for neutrino oscillations. Until
now no fully convincing sign of new physics has been detected,
except for the resonances of masses 1.8 TeV, 750 GeV and maybe 300
GeV, seen at LHC \ct{1,2,3,4}. These results caused a keen
interest in possibility of emergence of new physics only at very
high (Planck scale) energies. A largely explored scenario assumes
that new physics interactions appear only at the Planck scale:
\be M_{Pl}= 1.22\times 10^{19}\; {\rm GeV}. \lb{2} \ee
According to this scenario, we need the knowledge of the Higgs
effective potential $V_{eff}(\phi)$ up to very high values of
$\phi$. The loop corrections lead the $V_{eff}(\phi)$ to values of
$\phi$, which are much larger than $v$, where $v$ is the location
of the Electroweak (EW) vacuum. The effective Higgs potential
develops a new minimum at $v_2 \gg v$. The position of the second
minimum depends on the SM parameters, especially on the top and
Higgs masses, $M_t$ and $M_H$.

\section{Multiple Point Principle and the prediction of the top and Higgs
masses}

In general, a quantum field theory allows an existence of several
minima of the effective potential, which is a function of a scalar
field. If all vacua, corresponding to these minima, are
degenerate, having zero cosmological constants, then we can speak
about the existence of {\un{a multiple critical point (MCP)}} in
the phase diagram of theory \ct{5mp,6mp,7mp}).

In Ref.~\ct{5mp} Bennett and Nielsen postulated a Multiple Point Principle
(MPP) for many degenerate vacua.

See: {\bf Appendix A: Literature for MPP.}

This principle should solve the finetuning problem by actually
making a rule for finetuning. The Multiple Point Model (MPM) of
the Universe contains simply the SM itself up to the scale $\sim
10^{18}$ GeV. If the MPP is very accurate, we may have a new law
of Nature, that can help us to restrict coupling constants from
theoretical principles.

Assuming the existence of two degenerate vacua in the SM:
\begin{itemize}
\item the first Electroweak vacuum at v=246 GeV, and
\item the second Planck scale vacuum at $v_2\simeq 10^{18}$ GeV,
\end{itemize}
Froggatt and Nielsen predicted in Ref.~\ct{7mp} the top-quark and
Higgs boson masses:
 \be M_t = 173 \pm 5\; {\rm GeV }, \qquad M_H =
135 \pm 9\; {\rm GeV }. \lb{3} \ee
In Fig.~1 it is displayed the existence of the second
(non-standard) minimum of the effective potential in the pure SM
at the Planck scale.

The tree-level Higgs potential with the standard ``Electroweak
minimum" at $\phi_{min1} = v\approx 246$ GeV is given by:
\be V_1 = V {\rm (tree\; level)} = \lambda (\phi^2 - v)^2 + C_1. \lb{4} \ee
The new minimum at the Planck scale:
\be V_2 = V_{eff} {\rm (at\; Pl\; scale)} = \lambda_{run} (\phi^2 - v_2)^2 + C_2  \lb{5} \ee
can be higher or lower than the EW one, showing a stable EW vacuum
(in the first case), or metastable one (in the second case).

In accord with cosmological measurements, Froggatt and Nielsen
assumed that cosmological constants $C_1$ and $C_2$ for both vacua
are equal to zero (or approximately zero): $C_{1,2} = 0$, or
$C_{1,2}\approx 0$. This means that vacua $v=v_1$ and $v_2$ are
degenerate, or almost degenerate.

The following requirements must be satisfied in order that the effective potential
should have two degenerate minima:
\be  V_{eff}(\phi_{min1}^2) = V_{eff}(\phi_{min2}^2) = 0, \lb{6} \ee
and
\be  V'_{eff}(\phi_{min1}^2) = V'_{eff}(\phi_{min2}^2) = 0, \lb{7} \ee
where
\be V'(\phi^2) = \frac{\partial V}{\partial \phi^2}. \lb{8} \ee
As a result, Multiple Point Principle postulates: {\it there are
many vacua with the same energy density, or cosmological constant,
and all cosmological constants are zero, or approximately zero.}

If several vacua are degenerate, then the phase diagram of theory
contains {\un{a special point} --  the Multiple Critical Point
(MCP), at which the corresponding phases meet together.

Here it is useful to remind you a triple point of water analogy.

It is well known in the thermal physics that in the range of fixed
extensive quantities: volume, energy and a number of moles, the
degenerate phases of water (namely, ice, water and vapor,
presented in Fig.~2) exist on the phase diagram (P, T) of Fig.~3.

At the finetuned values of the variables -- pressure P and temperature T -- we have:
\be
T_c\approx 0.01^oC, \quad P_c \approx 4.58\quad {\rm{mm\; Hg}},
                                       \lb{9} \ee
giving the critical (triple) point $O$ shown in Fig.~3. This is a
{\un{triple point}} of water analogy.

The idea of the Multiple Point Principle has its origin from the
lattice investigations of gauge theories. In particular, Monte
Carlo simulations of $U(1)$-, $SU(2)$- and $SU(3)$-gauge theories
on lattice indicate the existence of the triple critical point.

\section{Cosmological Constant and Topological Structure of the Vacuum}

In the Einstein-Hilbert gravitational action:
\be
 S = \frac 1{8\pi G_N}\int_{\mathfrak M}d^4x \left(\frac R2 - \Lambda\right)
                                               \lb{10} \ee
(here $G_N$ is the Newton's gravitational constant),
Dark Energy (DE) -- vacuum energy density of our Universe --
is related with a cosmological constant $\Lambda$ by the following way:
\be       \rho_{DE} = \rho_{vac} = {(M_{Pl}^{red.})}^2\Lambda.  \lb{11} \ee
Here  $M^{red}_{Pl}$ is the reduced Planck mass:
\be M^{red}_{Pl}\simeq 2.43\times 10^{18}\; {\rm{ GeV}}. \lb{12c} \ee
Cosmological measurements gives:
\be \rho_{DE}\simeq (2\times 10^{-3}\; {\rm{eV}})^4,    \lb{13c} \ee
that means a tiny value of the cosmological constant:
\be  \Lambda \simeq 10^{-84}\; {\rm{GeV}}^4.   \lb{14c} \ee
By this reason, Bennett, Froggatt and Nielsen considered only zero, or
almost zero, cosmological constants for all vacua, existing in our Universe.

\subsection{Sidharth's theory of cosmological constant (Dark Energy)}

In 1997 year Sidharth suggested a model, in which the Universe
would be accelerating, driven by the so called Dark Energy,
corresponding to the extremely small cosmological constant
\ct{S1,S2}.

In 1998 year S. Perlmutter, B. Schmidt and A. Riess \ct{5} were
awarded by the Nobel Prize for discovery of the Universe
accelerating expansion.

We see that in papers \ct{S1,S2}:
\begin{enumerate}
\item Sidharth predicted a tiny value of the cosmological constant:
\be \Lambda \sim H_0^2,  \lb{15c} \ee
where $H_0$ is the Hubble rate in the early Universe;
\item Sidharth predicted that a Dark Energy (DE) density is very small:
\be \simeq 10^{-12}\; {\rm{eV}}^4 = 10^{-48}\; {\rm{GeV}}^4; \lb{16c} \ee
\item Sidharth predicted that a very small DE density provides an
accelerating expansion of our Universe after the Big Bang.
\end{enumerate}
Sidharth proceeded from the following points of view \ct{S3}:
Modern Quantum Gravity (Loop Quantum Gravity, etc.,) deal with a
non-differentiable space-time manifold. In such an approach, there
exists a minimal space-time cut off, which leads to the
non-commutative geometry, a feature shared by the Fuzzy
Space-Time also.

See: {\bf Appendix B. Non-commutativity, the main references.}

Following the book \ct{S3}, let us consider:
\begin{itemize}
\item $R_{un}$  --  the radius of the Universe $\sim 10^{28}$ cm,
\item $(T_{un})$  -- the age of the Universe,
\item $N_{un}$  --  the number of elementary particles in the Universe $(N_{un} \sim 10^{80})$,
\item $l$  -- the Compton wavelength of the typical elementary particle with mass $m$,\\
$(l = \hbar c/m)$ ($l\sim 10^{-10}$ cm for electron).
\end{itemize}
Then in a random walk, the average distance $l$ between particles is
\be l = R /\sqrt{N}, \lb{1s} \ee
and
\be
T_{un} = \sqrt{N_{un}} \tau ,\lb{2s}
\ee
where $\tau$ is a minimal time interval (chronon).

If we imagine that the Universe is a collection of the Planck mass oscillators,
then the number of these oscillators is:
\be
        {N}_{un}^{Pl}\sim 10^{120}. \lb{3s}
\ee
{\bf If the space-time is fuzzy, non-differentiable, then it has
to be described by a non-commutative geometry} with the
coordinates obeying the following commutation relations:
\be [dx^\mu , dx^\nu] \approx \beta^{\mu \nu} l^2 \ne 0.\lb{4s}
\ee
Eq.~(\ref{4s}) is true for any minimal cut off $l$.

Previously the following commutation relation was considered by
H.S. Snyder \ct{Sny}:
\be [x,p] = \hbar \left[1 + \left(\frac{l}{\hbar}\right)^2 p^2\right],\; etc., \lb{5s}
\ee
which shows that effectively 4-momentum $p$ is replaced by
\be p \to p \left(1 + \frac{l^2}{\hbar^2} p^2\right)^{-1}. \lb{6s} \ee
Then the energy-momentum formula now becomes as:
\be E^2 = m^2 + p^2 \left(1 + \frac{l^2}{\hbar^2} p^2\right)^{-2}, \lb{7s} \ee
or
\be E^2 \approx m^2 + p^2 - \gamma \frac{l^2}{\hbar^2} p^4, \lb{8s}
\ee
where $\gamma \sim 2$.

In such a theory the usual energy momentum dispersion relations are
modified \ct{S4}.

In the above equations $l$ stands for a minimal (fundamental)
length, which could be the Planck length, or for more generally -
Compton wavelength. It is necessary to comment that if we neglect
order of $l^2$ terms, then we return to the usual quantum theory.

Writing Eq.~(\ref{8s}) as
\be E = E' - E'',  \lb{9s} \ee
where $E'$ is the usual (old) expression for energy, and $E''$ is
the new additional term in modification. $E''$ can be easily verified as
\be E'' = mc^2. \lb{10s} \ee
In Eq.~(\ref{10s}) the mass $m$ is the mass of the field of bosons.
Furthermore it was proved, that (\ref{9s}) is valid only for boson fields,
whereas for fermions the extra term comes with a positive sign.
In general, we can write:
\be E = E' + E'' , \lb{11s} \ee
where $E'' = -m_b c^2$ -- for boson fields, and $E'' = + m_f c^2$
-- for fermion fields (with mass $m_b$, $m_f$, respectively).
These formulas help to identify the DE density, what was first
realized by B.G. Sidharth in Ref.~\ct{S2}.

DE density is the density of the quantum vacuum energy of the
Universe. Quantum vacuum, described by {\bf Zero Point Fields
(ZPF)} contributions, is the lowest state of any Quantum Field
Theory (QFT), and due to the Heisenberg's principle has an
infinite value, which is ``renormalizable".

As it was pointed out in Refs.~\ct{1Z,S5} that quantum vacuum of
the Universe can be a source of cosmic repulsion. However, a
difficulty in this approach has been that the value of the
cosmological constant turns out to be huge, far beyond what is
observed by astrophysical measurements. This has been called ``the
cosmological constant problem" \ct{1W}.

Using the non-commutative theory of the discrete space-time, B.G.
Sidharth predicted in \ct{S5} the value of cosmological constant
$\Lambda$:
\be \Lambda  \simeq H_0^2, \lb{14s} \ee
where $H_0$ is the Hubble rate:
\be   H_0 \simeq 1.5\times 10^{-42}\; {\rm{GeV}}. \lb{15s} \ee

\subsection{What is the  Universe vacuum?}

It is well known that in the early Universe topological defects
may be created in the vacuum during the vacuum phase transitions
\ct{1K,1BV}. It is thought that the early Universe underwent a
series of phase transitions, each one spontaneously breaking some
symmetry in particle physics and giving rise to topological
defects of some kind, which in many cases can play an essential
role throughout the subsequent evolution of the Universe.

In the context of the General Relativity, Barriola and Vilenkin
\ct{1BV} studied the gravitational effects of a global monopole as
a spherically symmetric topological defect. It was found that the
gravitational effect of global monopole is repulsive in nature.
Thus, one may expect that the global monopole and cosmological
constants are connected through their common manifestation as the
origin of repulsive gravity. Moreover, both cosmological constant
and vacuum expectation value are connected while the vacuum
expectation value is connected to the topological defect. All
these points lead us to a simple conjecture: There must be a
common connection among them, namely, the cosmological constant,
the global monopole (topological defect) and the vacuum
expectation value.

Remark: In the systematic phase of the early Universe, topological
defects were absent.

During the expansion of the early Universe, after the Planck era,
different phase transitions resulted in to the formation of the various
kind of much discussed topological defects like monopoles (point
defects), cosmic strings (line defects) and domain walls (sheet
defects). The topology of the vacuum manifold dictates the nature of
these topological defects. These topological defects appeared due to
the breakdown of local or global gauge symmetries.
In Ref.~\ct{GuRa} it was studied the gravitational field, produced by a spherically
symmetric ``hedgehog" configuration in scalar field theories with a
global $SO(3)$ symmetry.

For isovector scalar:
\be \Phi = (\Phi_1,\; \Phi_2,\; \Phi_3,\; ) \lb{16S} \ee
this solution is pointing radially, what means that $\Phi$ is parallel to $\hat{r}$ ,
the unit vector in the radial direction. The started Lagrangian of this theory is:
\be L = \frac 12 \partial_{\mu}\Phi \cdot \partial_{\nu}\Phi g^{\mu\nu}
+ \lambda(\Phi \cdot \Phi - v^2)^2.
                                                 \lb{17S} \ee
If  $\Phi$  is constraint as $\Phi \cdot \Phi = v^2$
(for example, at $|\Phi| \to \infty$), then the Lagrangian is:
\be L = \frac 12 \partial_{\mu}\Phi \cdot \partial_{\nu}\Phi
g^{\mu\nu}.  \lb{18S} \ee Topological structures in fields are as
important as the fields themselves. In Ref.~\ct{BCK} the
gauge-invariant hedgehog-like structures in the Wilson loops were
investigated in the $SU(2)$ Yang-Mills theory. In this model the
triplet Higgs field $\Phi \equiv \frac 12 \Phi^a \sigma^a$ ($a$ =
1, 2, 3) vanishes at the center of the monopole $x = x_0$:
\be \Phi(x_0) = 0 \ee
and has a generic hedgehog structure in the spatial vicinity of
this monopole.

Recently in arXiv appeared the investigation \ct{Lus} (see also
Ref.~\ct{Deli}).

In Refs.~\ct{Lus,Deli} the authors obtained a solution for a
black-hole in a region that contains a global monopole in the
framework of the $f(R)$ gravity, where $f(R)$ is a function of the
Ricci scalar $R$. Near the Planck scale they considered the
following action:
\be  S = \frac 1{\kappa^2} \int d^4x \left(f(R) + \frac 12
D_{\mu}^b(\phi^b)^\dagger D^{\mu}_a\phi^a - \frac 14 (|\phi^a|^2 -
v^2)^2 +...\right), \lb{19s} \ee
where $\kappa^2 = 8\pi G_N$, $G_N$ is the Newton's gravitational
constant, $\Phi^a$ is the Higgs triplet field ($a$ = 1, 2, 3),
$\lambda$ is the Higgs self-interaction coupling, and $v$ (which
here is $v_2$) is the vacuum expectation value (VEV) of $\Phi$ at
the Planck scale:
\be v= v_2=\langle\Phi_{min2}\rangle\sim 10^{18}\; {\rm{GeV}}. \lb{20s} \ee
Here $D_{\mu}$ is a covariant derivative:
\be D_{\mu}^a = \partial_{\mu} + i\omega_{\mu}^{a} + iW_{\mu}^{a}, \lb{21s} \ee
where $\omega_{\mu}^{a}$ is the gravitational spin-connection, and
$W_{\mu}^{a}$ is the $SU(2)$ gauge field.

Considering the time independent metric with spherical symmetry in $(3+1)$ dimensions:
\be ds^2 = B(r)dt^2 - A(r)dr^2 - r^2\big(d\theta^2 + sin(2\theta)d\varphi^2\big), \lb{22s} \ee
the authors of Refs.~\ct{Lus,Deli} obtained a monopole configuration, which is described as:
\be \phi^a = v_2 h(r)\frac{x^a}{r}, \lb{23s} \ee
where $a$ = 1, 2, 3 and $x^ax^a = r^2$. This is ``a hedgehog"
solution by Alexander Polyakov's terminology.

In the flat space the hedgehog core has the size:
\be   \delta \sim \frac 1{\sqrt {\lambda} v}, \lb{24s} \ee
and the mass:
\be M_{core} \sim \frac {v}{\sqrt {\lambda}},  \lb{25s} \ee
which is:
\be M_{BH} \sim  M_{Pl} \sim 10^{-5}\; {\rm{gms}},   \lb{26s} \ee
or
\be   M_{BH} \sim 10^{18}\; {\rm{GeV}}. \lb{27s} \ee
This is a black-hole solution, which corresponds to a global
monopole that has been \un{swallowed} by a black-hole.

Now we see, that the Planck scale Universe is described by a
non-differentiable space-time: by a foam of black-holes, having
lattice-like structure, in which sites are black-holes with the
``hedgehog" monopoles inside them.

Global monopole is a heavy object formed as a result of
gauge-symmetry breaking in the phase transition of an isoscalar
triplet $\Phi^a$ system. The black-holes-monopoles-hedgehogs are
similar to elementary particles, because a major part of their
energy is concentrated in a small region near the monopole core.
In the Guendelman-Rabinowitz theory \ct{GuRa}, a gravitational
effect similar to hedgehogs can be generated by a set of cosmic
strings in a spherically symmetric configuration, which can be
referred to as a ``string hedgehog". The authors investigated the
evolution of bubbles separating two phases: one being the ``false
vacuum" (Planck scale vacuum) and the other  the ``true vacuum"
(EW-scale vacuum). The presence of the hedgehogs, called
``defects", is responsible for the destabilization of a false
vacuum. Decay of a false vacuum is accompanied by the growth of
the bubbles of a true vacuum. Guendelman and Rabinowitz also
allowed a possibility to consider an arbitrary domain wall between
two phases. During the inflation domain wall annihilates,
producing gravitational waves and a lot of SM particles, having
masses.

The non-commutative contribution of the black-holes of the Planck scale vacuum
compensates the contribution of the Zero Point Fields
and the cosmological constant of the Planck scale phase is:
\be  \Lambda ({\rm at\; Pl.\; scale}) = \Lambda_{ZPF}({\rm at\; Pl.\; scale})
- \Lambda_{BH} = 0. \lb{28s} \ee
That is, the phase with the VEV $v=v_2$ has zero cosmological constant.

By cosmological theory, the Universe exists in the Planck scale
phase for extremely short time. By this reason, the Planck scale
phase was called ``the false vacuum". After the next phase
transition, the Universe begins its evolution toward the second,
Electroweak (EW) phase. Here the Universe underwent the inflation,
which led to the phase having the VEV:
\be v = v_1\approx 246\; {\rm{GeV}}. \lb{29s} \ee
The EW (``true") vacuum with the VEV $v\approx 246$ GeV is the
vacuum, in which we live.

\section{Phase transition(s) in the Universe}

Now it is useful to understand the effects of the finite temperature on
the Higgs mechanism (see \ct{MSher}).

At some finite temperature which is called the critical
temperature $T_c$, a system exhibits a spontaneous symmetry
breaking. A ferromagnet is an example of spontaneously broken
symmetry. In this theory the equations of motion are rotationally
symmetric, but the ground state of a ferromagnet has a preferred
direction.

In the Landau and Ginzburg theory \ct{GL}, the free energy of an
isotropic ferromagnet is:
\be  F = \frac 12 \alpha |M|^2 + \frac 14\beta |M|^4, \lb{1pt} \ee
where $\alpha$ is positive and $M$ is the magnetization, $\alpha$
has a temperature dependence, and near the critical point it is
given by $\alpha= \alpha_0(T - T_c)$. Thus, for temperatures below
the critical temperature,  $\alpha$ is negative, and the vacuum
value of $|M|$ is nonzero. For temperatures above the critical
temperature, $\alpha$  is positive, and the magnetization
vanishes. This is what one intuitively expects: at high
temperatures the kinetic energy of the atoms is much greater than
the spin exchange interaction energy, thus the average
magnetization should vanish. Therefore, at high temperatures the
rotational $O(3)$ symmetry of a ferromagnet is restored.

The spontaneous symmetry breakdown of a gauge theory also vanishes
at high temperature, and the gauge symmetry is restored. Kirzhnits
\ct{Kir} and Linde \ct{1KL}  were first who considered the analogy
between the Higgs mechanism and superconductivity, and argued that
the Higgs field condensate disappears at high temperatures,
leading to symmetry restoration. As a result, in the Higgs model
at high temperatures, all fermions and vector bosons are massless.
These conclusions were confirmed, and the critical temperature was
estimated in Refs.~\ct{DJ,SW,2KL}. See also the review article by
A.~Linde \ct{Lin}.

Let us consider now the phase transition from a false vacuum to a true vacuum.
At the early stage the Universe was very hot, but then it began to
cool down. Black-holes-monopoles (as bubbles of the vapor in the
boiling water) began to disappear. The temperature dependent part of the energy density died away.
In that case, only the vacuum energy will survive. Since this is a
constant, the Universe expands exponentially, and an exponentially
expanding Universe leads to the inflation (see \ct{Guth,Star,1Lin}, etc.).

During the inflation the triplet Higgs field, $\phi^a$, $a$ = 1, 2, 3, decays into
the Higgs doublet fields of $SU(2)_L$, $\Phi$. Here we follow to the
Gravi-Weak Unification theory of Ref.~\ct{DLT}, and finally we have the Standard Model
Lagrangian with gravity:
\be    \frac 12 f(R) - \Lambda + \frac 12 D_{\mu}\Phi^\dagger D^{\mu}\Phi
-\frac 14\lambda (|\Phi|^2 - v^2)^2 +  L_{matter} + L_{gauge} + L_{Yuk},
                              \lb{2pt} \ee
where $\Lambda$ is a cosmological constant, and
\be   D_{\mu} = \partial_{\mu} + i\omega_{\mu}^{i}\varrho_i + iW_{\mu}^{i}\tau_i +iA_{\mu}
                                 \lb{3pt} \ee
is a covariant derivative.

$L_{matter}$, $L_{gauge}$ and $L_{Yuk}$ are respectively the
matter fields Lagrangian (including quarks with flavors $f$ and
leptons $e,\; \nu $), the gauge fields Lagrangian (including
gluons $G_{\mu}$, vector bosons $W_{\mu}$ and the electromagnetic
field $A_{\mu}$), and the Yukawa couplings Lagrangian of type $Y_f
\bar \psi_f\psi_f\Phi$. Here: \be {{\mathfrak sl}(2,C)}^{(grav)}:
\quad \{\rho_i\}=\{\sigma_i\otimes 1_2\},\ee and \be {{\mathfrak
su}(2)}^{(weak)}:\quad \{\tau_i\}=\{1_2\otimes \sigma_i\}.\ee The
Electroweak vacuum has the Higgs field's VEV: $v\approx 246$ GeV.

While the Universe was being in the false vacuum and expanding
exponentially, so it was cooling exponentially. This scenario was
called {\bf supercooling in the false vacuum.} When the
temperature reached the critical value $T_c$, the Higgs mechanism
of the SM created a new condensate $\Phi_{min 1}$, and the vacuum
became similar to superconductor, in which the topological defects
are the closed magnetic vortices. The energy of black-holes is
released as particles, which were created during the radiation era
of the Universe, and all these particles (quarks, leptons, vector
bosons) acquired their masses through the Yukawa coupling
mechanism $Y_f \bar \psi_f\psi_f\Phi$.

The electroweak spontaneous breakdown of symmetry $SU(2)_L\times
U(1)_Y \to U(1)_{el.mag}$ leads to the creation of the topological
defects in the EW vacuum. They are the Abrikosov-Nielsen-Olesen
closed magnetic vortices of the Abelian Higgs model \ct{Ab,NO}.
Then the electroweak vacuum again presents the non-differentiable
manifold, and again we have to consider the non-commutative
geometry, in accordance with the Sidharth's theory of the
vacuum.

However, here we have fermions, which have a mass, therefore
Compton wavelength, $\lambda=\hbar/mc$, and according to the
Sidharth's theory of the cosmological constant, we have in the
EW-vacuum lattice-like structure of bosons and fermions with
lattice parameter ``$l$" equal to the Compton wavelength: $l =
\hbar/mc$.

Taking into account the relation between the vacuum energy
density, $\rho_{vac}$, and the cosmological constant $\Lambda$:
\be  \rho_{vac} = \rho_{DE} = {M_{Pl}^{red.}}^2 \Lambda, \lb{4pt}
\ee
we easily see that in the Planck scale vacuum (with the VEV $v_2 \sim 10^{18}$ GeV)
we have:
\be   \rho_{vac}({\rm at\; Planck\; scale}) =  \rho_{ZPF}({\rm
at\; Planck\; scale}) - \rho_{black\; holes}^{(NC)}\approx 0,
\lb{5pt} \ee
and
\be   \rho_{vac}({\rm at\; EW\; scale}) =  \rho_{ZPF}({\rm at\; EW\; scale})
- \rho_{vortex\; contr.}^{(NC)} -\rho_{boson\; fields}^{(NC)}
   + \rho_{fermion\; fields}^{(NC)} \approx 0, \lb{6pt} \ee
In the above equations ``NC" means the ``non-commutativity" and
``ZPF" means ``zero point fields".

Here it is necessary to comment that the correctness of an
assessment of the non-commutative theory is well known in the
paper Ref.~\ct{SDR}. Here we emphasize that, due to the energy
conservation law, the vacuum density before the phase transition
at the critical temperature $T_c$ is equal to the vacuum density
after the phase transition, that is:
\be  \rho_{vac}({\rm at\; Planck\; scale}) = \rho_{vac}({\rm at\; EW\; scale}).
                                                  \lb{7pt} \ee
The analogous link between the Planck scale phase and EW phase was
considered in the paper \ct{S6}. It was shown that the vacuum
energy density (DE) is described by the different contributions to
the Planck and EW scale phases. This difference is a result of the
phase transition. However, the vacuum energy densities (DE) of
both vacua are equal, and we have a link between gravitation and
electromagnetism via the Dark Energy, established in Ref.~\ct{S6}.
According to the last equation (\ref{7pt}), we see that if
$\rho_{vac}$(at Planck scale) is almost zero, then $\rho_{vac}$(at
EW scale) also is almost zero, and we have a triumph of the
Multiple Point Principle!

A general theory of the phase transition from the one type lattice
structure to the another type, in particular, from the Planck
scale lattice with sites $\phi_p$ to the Compton scale lattice
with sites $\phi_c$, was developed in Refs.~\ct{S7,S8}. Previously
it has been proved that the phase transition from the Planck scale
phase to the Compton scale (EW) phase is similar to the
Landau-Ginzburg phase transition \ct{S9}. In these investigations
it has been substantiated that the 2D universe undergoes a phase
transition from the Planck phase to the Compton phase in analogy
with the ferromagnetic case.

This result should also be hold in the case of a 3D universe.

Here it is worth to specify the concept ``Compton phase" entered by
Sidharth. Taking into account Sidharth's previous works
\ct{S9,S10}, we have, in analogy with a coherence parameter $\xi$
of the Ginzburg-Landau theory \ct{GL}, the following coherence
parameter:
\be \xi = \frac{hc}{\Delta},\quad \Delta = mc^2, \quad \xi = \frac {2\pi \hbar}{mc}
= 2\pi l_c, \lb{8pt}, \ee
where $l_c = \frac {\hbar}{mc}$ is the Compton length of a particle having mass $m$.

If we consider the Higgs particle (with mass $m_H$), then we have
it's Compton length: \be l_H = \frac {\hbar}{m_Hc}, \ee i.e. the
coherence parameter of the phase under consideration is the
Compton length of the Higgs boson. In general, we can say: The
Compton length is the fundamental aspect of the Compton phase,
which is a synonym to the Electroweak phase, the current phase of
the Universe. Thus, B.G. Sidharth explained in his investigations,
why the Compton scale plays such a rudimentary role in all
phenomena of the quantum physics. The Compton scale gives the
description of an accelerating Universe with a small positive
cosmological constant \ct{S2}.

In the paper \ct{SDR1} it was obtained that the Compton scale
gives the correction to the electron anomalous gyromagnetic ratio
$g = 2$, what also was considered by J.~Schwinger from the quantum
field theoretical point of view.

The papers \ct{SDR2} and \ct{DS} were devoted to the Lamb shift as
a phenomenon that can be attributed only to the Compton scale and
to the non-commutative nature of the space-time.

\section{Vacuum stability and the Multiple Point Principle}

The vacuum stability problem in the Standard Model has a long
history:

See: {\bf Appendix C and references in \ct{LND}.}

If $\Lambda_{EW} > \Lambda_{Pl}$ what means:
\be \rho_{vac}({\rm at\; EW\; scale}) > \rho_{vac}({\rm at\;
Planck\; scale}),     \lb{1vs}, \ee
than our vacuum is not stable, it decays! And the MPP is not
exact.

For energies higher than EW scale the analysis of the vacuum
stability is reduced to the study of the renormalization group evolution
of the Higgs quartic coupling $\lambda$ (see \ct{MSher}).

The Froggatt-Nielsen's prediction for the mass of the Higgs boson
$M_t = 173 \pm 5$ GeV; $M_H = 135 \pm 9$ GeV was improved  in
Ref.~\ct{Deg} by the calculation of the two-loop radiative
corrections to the effective Higgs potential.

The prediction of Higgs mass $129.4 \pm 1.8$ GeV is closed the
value $M_H \simeq 125.7$ GeV observed at the LHC. The authors of
Ref.~\ct{But} have shown that the most interesting aspect of the
measured value of $M_H$ is its near-criticality. They extrapolated
the SM parameters up to the high (Planck scale) energies with full
three-loop NNLO RGE precision.

The main result of the investigation of \ct{Deg}  is: The observed
Higgs mass $M_H = 125.66 \pm 0.34$ GeV at LHC leads to the
negative value of the Higgs quartic coupling $\lambda$ at some
energy scale below the Planck scale, making the Higgs potential
unstable or metastable. For the vacuum stability investigation a
highly precise analysis is quite necessary.

With the inclusion of the three-loop RG equations in \ct{But} and
two-loop matching conditions in \ct{Deg}, the instability scale
occurs at $10^{11}$ GeV, well below the Planck scale. This means
that at that scale the effective potential starts to be negative,
and a new minimum can appear with negative cosmological constant.
According to these investigations, the experimental value of the
Higgs mass gives scenarios, which are at the borderline between
the absolute stability and metastability. The measured value of
$M_H$ puts the Standard Model in the so-called near-critical
position. Using the present experimental uncertainties on the SM
parameters (mostly the top-quark mass) it is conclusively
impossible to establish the fate of the EW vacuum, although
metastability is preferred. Thus, the careful evaluation of the
Higgs effective potential by Ref.~\ct{Deg}, combined with the
experimentally measured Higgs boson mass in the pure SM, leads to
the energy density getting negative for high values of the Higgs
field, what means that the minimum of the effective potential at
$10^{18}$ GeV (if it exists) has a negative energy density.
Therefore, formally the vacuum, in which we live, is unstable
although it is in reality just metastable with an enormously long
life-time. However, only this unstable vacuum corresponding to the
experimental Higgs mass of $125.66 \pm 0.34$ GeV is indeed very
close to the Higgs mass $129.4 \pm 1.8$ GeV obtained by Degrassi
et al. in Ref.~\ct{Deg}. The last value makes the $10^{18}$ GeV
Higgs field vacuum be degenerate with the Electroweak one. In this
sense, Nature has chosen parameters very close to ones predicted
by the Multiple Point Principle.

\subsection{Could the Multiple Point Principle be exact due to
corrections from the new bound state 6t + 6anti-t ?}

See: {\bf Appendix D. Theory of the new bound state 6t + 6anti-t,
the main references.}

The purpose of the articles of Refs.~\ct{LND,DLN} is to estimate
the correction from the NBS $6t + 6\bar t$ to the Higgs mass
$129.4 \pm 1.8$ GeV obtained by Degrassi et al. in Ref.~\ct{Deg}.
This is actually can be done by identifying a barely significant
peak obtained at the LHC Run2 with proton-proton collisions at
energy $\sqrt{s} = 13$ TeV in the LHC-experiments
\ct{LHC-1,LHC-2,LHC-3}.

\subsubsection{LHC: Search for the resonances in pp collision data
at $\sqrt s = 13$ TeV}

Recently the ATLAS and CMS collaborations \ct{1,2,3,4} have
presented the first data obtained at the LHC Run 2 with pp
collisions at energy $\sqrt s = 13 $ TeV. Fig.~4 presents
searches for a new physics in high mass diphoton events in
proton-proton collisions at 13 TeV. ATLAS and CMS Collaborations
show a new resonance in the diphoton distribution at the invariant
mass of 750-760 GeV.

The ATLAS collaboration  claims an excess in the distribution of
events containing two photons, at the diphoton invariant mass $M
\approx 750$ GeV with $3.9\sigma$ statistical significance. The
ATLAS excess consists of about 14 events suggesting a best-fit
width $\Gamma$ of about 45 GeV with $\Gamma/M \approx 0.06$.

See: {\bf Appendix E. Resonance 750 GeV.}

ATLAS \ct{1,2,3,4} collaboration  presents searches for resonant
and non-resonant Higgs boson pair production in proton-proton
collision data at $\sqrt s = 8$ TeV generated by the LHC and
recorded by the ATLAS detector in 2012 (see Fig.~5). In the
search for a narrow resonance decaying to a pair of Higgs bosons,
the expected exclusion on the production cross section falls from
1.7 pb for a resonance at 260 GeV to 0.7 pb at 500 GeV. It is not
excluded that then results show: a resonance with
mass $\approx 300-350$ GeV.

If the observed diphoton excess indeed corresponds to the decay of
a hitherto unknown particle then this will be the first
confirmation of new physics beyond the SM. If the observed excess
is due to a resonance it has to be a boson and it cannot be a
spin-1 particle \ct{L,Y}. This leaves the possibility of it being
either a spin-0 or spin-2 particle \ct{MSS}. If it is indeed a new
particle, then one must wonder what kind of new physics
incorporates it.

In previous Ref.~\ct{FNL} we have speculated that $6t + 6\bar t$
quarks should be so strongly bound that these bound states would
effectively function at low energies as elementary particles and
can be added into loop calculations as new elementary particles or
resonances. The exceptional smallness of the mass $m_S$ of the new
bound state particle $S$:
\be    m_S \ll 12M_t \lb{2vs}, \ee
is in fact a consequence of the degeneracy of the vacua, and thus
of the Multiple Point Principle.

Run 2 LHC data show hints of a new resonance in the diphoton
distribution at an invariant mass of 750 GeV. We identify this peak with
our NBS $6t + 6\bar t$. It means that taking into account the contribution of
the LHC resonance with mass 750 GeV as an our bound state $S$
during the calculations of the correction to the predicted Higgs mass,
we must obtain a new result for the vacuum stability and MPP.

\section{The Higgs effective potential}

A theory of a single scalar field (see Ref.~\ct{MSher}) is given by the
effective potential $V_{eff}(\phi_c)$, which is a function of the classical field $\phi^c$.
In the loop expansion this $V_{eff}$ is given by:
\be V_{eff} = V^{(0)} + \sum_{n=1} V^{(n)}, \lb{1a} \ee
where $V^{(0)}$ is the tree level potential of the SM.

The Higgs mechanism is the simplest mechanism leading to the
spontaneous symmetry breaking of a gauge theory. In the SM the breaking
\be
 SU(2)_L\times U(1)_Y \to U(1)_{em}, \lb{2a}
\ee
achieved by the Higgs mechanism, gives masses to the Higgs and
gauge bosons, also to fermions with flavor $f$.

With one Higgs doublet of $SU(2)_L$, we have the following
tree level Higgs potential:
\be V^{(0)} = - m^2 \Phi^{+}\Phi + \lambda(\Phi^{+}\Phi )^2.
 \lb{3a} \ee
The vacuum expectation value of $\Phi$ is:
\be
 \left\langle \Phi\right\rangle = \frac{1}{\sqrt 2}\left(
 \ba{c}
 0\\
 v
 \ea
 \right), \lb{4a} \ee
where
\be
 v = \sqrt{\frac{m^2}{\lambda}}\approx 246\; {\mbox{GeV}}.
 \lb{5a} \ee
Introducing a four-component real field $\phi$ by
\be
 \Phi^{+}\Phi = \frac{1}{2}\phi^2, \lb{6a} \ee
where
\be
 \phi^2 = \sum_{i=1}^4 \phi_i^2, \lb{7a} \ee
we have the following tree level potential:
\be
 V^{(0)} = - \frac{1}{2} m^2 \phi^2 + \frac{1}{4} \lambda
 \phi^4. \lb{8a} \ee
As is well-known, this tree-level potential gives the masses of the
gauge bosons $W$ and $Z$, fermions with flavor $f$ and the physical Higgs
boson $H$:
\bea
M_W^2 &=& \frac{1}{4} g^2 v^2, \lb{9a}\\
M_Z^2 &=& \frac{1}{4} \left(g^2 + g'^2\right) v^2, \lb{10a}\\
M_f &=& \frac{1}{\sqrt 2} g_f v, \lb{11a}\\
M_H^2 &=& \lambda v^2,            \lb{12a}
\eea
where $g_f$ is the Yukawa couplings of fermion with the flavor
$f$; $ g,\; g'$ are respectively $SU(2)_L$ and $U(1)_Y$ coupling constants.

\section{Stability phase diagram}

We have already seen that the new physics can come only at very
high (Planck scale) energies: $M_{Pl\,\,scale} \sim 10^{19}$ GeV.
The loop corrections lead the $V_{eff}(\phi)$ to the very large
(Planck scale) values of $\phi$, much larger than $v$, the
location of the EW vacuum. The effective Higgs potential develops
a new minimum at $v_2 \gg v$. The position of the second minimum
depends on the SM parameters, especially on the top and Higgs
masses, $M_t$ and $M_H$. It can be higher or lower than the EW
minimum, showing a stable EW vacuum (in the first case), or
metastable one (in the second case).

Considering the lifetime $\tau$ of the false vacuum (see
Ref.~\ct{BMS}) and comparing it with the age of the Universe
$T_U$, we see that, if $\tau$ is larger than $T_U$, then our
Universe will be sitting in the metastable vacuum, and we deal
with the scenario of metastability. The stability analysis is
presented by the stability diagram in the plane ($M_H,\; M_t$)
given by Fig.~6.

The stability line separates the stability and the metastability
regions, and corresponds to $M_t$ and $M_H$ obeying the condition
$V_{eff}(v) = V_{eff}(v_2)$. The instability line separates the
metastability and instability regions. It corresponds to $M_t$ and
$M_H$ for $\tau = T_U$. In the stability figure the black dot
indicates current experimental values $M_H\simeq 125.7$ GeV and
$M_t \simeq 173.34$ GeV: see Particle Data Group \ct{KAO}.

It lies inside the metastability region. The ellipses take into
account $1\sigma,\; 2\sigma$ and $3\sigma$, according to the
current experimental errors.

When the black dot sits on the stability line, then this case is
named ``critical", according to the MPP concept: then the running
quartic coupling $\lambda$ and the corresponding beta-function
vanish at the Planck scale $v_2$:
\be  \lambda\left(M_{Pl}\right) \sim 0 \quad {\rm{and}}\quad
\beta\left(\lambda\left(M_{Pl}\right)\right)\sim 0. \lb{13a} \ee
Stability phase diagram shows that the black dot, existing in the
metastability region, is close to the stability line, and this
``near-criticality" can be considered as the most important
information obtained for the Higgs boson.

\subsection{Two-loop corrections to the Higgs mass from the effective potential}

Still neglecting the new physics interactions at the Planck scale, we
can consider the Higgs effective potential $V_{eff}(\phi)$ for large values of $\phi$:
\be  V_{eff}(\phi)\simeq \frac 14 \lambda_{eff}(\phi) \phi^4.  \lb{14a} \ee
Here $V_{eff}(\phi)$ is the renormalization group improved (RGE)
Higgs potential (see \ct{MSher}), and $\lambda_{eff}(\phi)$
depends on $\phi$ as the running quartic coupling $\lambda(\mu)$
depends on the running scale $\mu$. Then we have the one-loop,
two-loops or three-loops expressions for $V_{eff}$ . The
corresponding up to date Next-to-Next-to-Leading-Order (NNLO)
results were published by Degrassi et al.~\ct{Deg} and Buttazzo et
al.~\ct{But}.

The relation between $\lambda$ and the Higgs mass is:
\be \lambda(\mu) = \frac{G_F}{\sqrt 2} M_H^2 + \Delta
\lambda(\mu), \lb{18a} \ee
where $G_F$ is the Fermi coupling. Here $\Delta \lambda(\mu)$
denotes corrections arising beyond the tree level potential.
Computing $\Delta \lambda(\mu)$ at the one-loop level, using the
two-loop beta functions for all the Standard Model couplings,
Degrassi et al.~\ct{Deg} obtained the first complete NNLO
evaluation of $\Delta \lambda(\mu)$. In the RGE figure blue lines
(thick and dashed) present the RG evolution of $\lambda(\mu)$ for
current experimental values $M_H \simeq 125.7$ GeV and $M_t \simeq
173.34$ GeV, and for $\alpha_s$ given by $\pm 3\sigma$.

The thick blue line corresponds to the central value of $\alpha_s
= 0.1184$ and dashed blue lines correspond to errors of $\alpha_s$
equal to $\pm 0.0007$. Absolute stability of the Higgs potential
is excluded by the investigation \ct{Deg} at 98\% C.L. for $M_H <
126$ GeV. In figure we see that asymptotically $\lambda(\mu)$ does
not reach zero, but approaches to the negative value, indicating
the metastability of the EW vacuum:
\be \lambda \to - (0.01 \pm 0.002), \lb{19a} \ee
According to Degrassi et al.~\ct{Deg}, the stability line is the
red thick line in the figure, and corresponds to: $M_H = 129.4 \pm
1.8$ GeV. The aim of Refs.~\ct{LND,DLN} is to show that the
stability line could correspond to the current experimental values
of the SM parameters, with $M_H = 125.7$ GeV given by LHC,
provided we include a correction caused by the newly found at LHC
resonance, which is identified as the bound state of our $6t +
6\bar t$.

\subsection{The effect from the new bound states 6t + 6anti-t on the
measured Higgs mass}

In Ref.~\ct{FNL} was first assumed that:

\begin{enumerate}
\item there exists $1S$-bound state $6t + 6\bar t$ -- scalar particle and color
singlet;
\item that the forces responsible for the formation of these bound states
originate from the virtual exchanges of the Higgs bosons between
top(anti-top)-quarks;
\item that these forces are so strong that they almost compensate the
mass of 12 top(anti-top)-quarks contained in these bound states.
\end{enumerate}

The explanation of the stability of the bound state $6t + 6\bar t$ is
given by the Pauli principle:
top-quark has two spin and three color degrees of freedom (total 6).
By this reason, 6 quarks have the maximal binding energy, and 6 pairs
of $t\bar t$ in $1S$-wave state create a long lived (almost stable) colorless
scalar bound state $S$. One could even suspect that not only this most
strongly bound state $S$ of $6t + 6\bar t$, but also some excited states exist,
and a new bound state $6t + 5\bar t$, which is a fermion similar to the quark
of the 4th generation.

These bound states are held together by exchange of the Higgs and
gluons between the top-quarks and anti-top-quarks as well as between
top and top and between anti-top and anti-top.
The Higgs field causes attraction between quark and quark as well as
between quark and anti-quark and between anti-quark and anti-quark,
so the more particles and/or anti-particles are being put together the
stronger they are bound.
But now for fermions as top-quarks, the Pauli principle prevents too
many constituents being possible in the lowest state of a Bohr atom
constructed from different top-quarks or anti-top-quarks surrounding
(as electrons in the atom) the ``whole system"  analogous to the
nucleus in the Bohr atom.

Because the quark has three color states and two spin states meaning
6 internal states there is in fact a shell (as in the nuclear physics) with
6 top-quarks and similarly one for 6 anti-top-quarks. Then we imagine
that in the most strongly bound state just this shell is filled and closed
for both top and anti-top. Like in nuclear physics where the closed shell
nuclei are the strongest bound, we consider this NBS $6t + 6\bar t$ as our
favorite candidate for the most strongly bound and thus the lightest
bound state $S$. Then we expect that our bound state $S$ is appreciably
lighter than its natural scale of 12 times the top mass, which is about
2 TeV. So the mass of our NBS $S$ should be small compared to 2 TeV.
Estimating different contributions of the bound state $S$, we have
considered the main Feynman diagrams correcting the effective Higgs
self-interaction coupling constant $\lambda(\mu)$. They are diagrams containing
the bound state $S$ in the loops.

\subsection{The effect from the new bound states $6t + 6\bar t$ on the
measured Higgs mass. The main diagrams correcting
the effective Higgs self-interaction coupling constant $\lambda$}

Now we have the following running $\lambda(\mu)$:
\be \lambda(\mu) = \frac{G_F}{\sqrt 2} M_H^2 + \delta
\lambda(\mu) + \Delta \lambda(\mu), \lb{19a} \ee
where the term $\delta\lambda(\mu)$ denotes the loop corrections
to the Higgs mass arising from the NBS, and the main contribution
to $\delta\lambda(\mu)$ is the term $\lambda_S$, which corresponds
to the contribution of the first Feynman diagram:
\be \delta \lambda(\mu) = \lambda_S + ... \lb{20a} \ee
The rest contributions of the Feynman diagrams are shown in
Ref.~\ct{LND}.

You can see the result of the corrections to the running $\lambda$ from the
bound state $S$ in the recent papers \ct{LND,DLN}.

The result is:
\be \lambda_S\approx \frac{1}{\pi^2}\left( \frac{6g_t}{b}\times
\frac{m_t}{m_S}\right)^4, \lb{21a} \ee
where $g_t$ is the experimentally found Yukawa coupling of top-quark
with the Higgs boson, $m_t$ and $m_S$ are masses of the top-quark and
$S$-bound state, respectively, and $b$ is a parameter, which determines
the radius $r_0$ of the bound state $S$:
\be r_0 = \frac {b}{m_t}. \lb{22a} \ee
As we see, the figure given by Degrassi et al.~\ct{Deg} showed that asymptotically
$\lambda(\mu)$ does not reach zero, but approaches to the negative value:
\be  \lambda \to -(0.01\pm 0.002),  \ee
indicating the metastability of the EW vacuum.

If any resonance gives the contribution:
\be   \lambda \to + 0.01, \ee
then this contribution transforms the metastable (blue) curve of
the stability diagram into the red curve, which is the borderline
of the stability (see Fig.~7).

Using the results obtained earlier in Ref.~\ct{FN},
we have calculated in Ref.~\ct{LND} the value of the $S$-bound state's radius:
\be      r_0 \approx \frac {2.34}{m_t}.   \lb{23a} \ee
Such radius of $S$ gives:
\be \lambda_S \simeq 0.009, \ee
or taking into account that the uncertainty coming from the
contributions of the rest Feynman diagrams can reach 25\%, we have
finally:
\be  \lambda_S \simeq 0.009 \pm 0.002. \ee
Just this result for radius provides the vacuum stability in the
Standard Model confirming the accuracy of the Multiple Point
Principle.

\section{Summary and Conclusions}

\begin{enumerate}
\item  We reviewed the Multiple Point Model (MPM) by D.L.~Bennett
and H.B.~Nielsen. We showed that the existence of two vacua into
the Standard Model: the first one  at the Electroweak scale ($v =
v_1 \approx 246$ GeV), and the second one  at the Planck scale
($v_2\sim 10^{18}$ GeV), was confirmed by calculations of the
Higgs effective potential in the two-loop and three-loop
approximations. The Froggatt-Nielsen's prediction of the top-quark
and Higgs masses was given in the assumption that there exist two
degenerate vacua in the Standard Model. This prediction was
improved by the next order calculations.

\item   Here we have reviewed the Sidharth's theory of the
cosmological constant  theory of the vacuum energy density of
our Universe, or Dark Energy. B.G. Sidharth was to show (in
1997) that the cosmological constant $\Lambda$ is extremely small:
$\Lambda\sim H_0^2,$ where $H_0$ is the Hubble rate, and the Dark Energy density
is very small ($\sim 10^{-48}$ ${\rm GeV}^4$), what provided the accelerating
expansion of our Universe after the Big Bang.

\item We considered the theory of the vacua of the Universe Planck
scale phase and Electroweak phase. Considering the topological
defects in these vacua, we have discussed that topological defects
of the Planck scale phase are black holes solutions, which
correspond to the ``hedgehog" monopole that has been ``swallowed"
by a black hole. It was suggested to consider the topological
defects in the Electroweak phase as Abrikosov-Nielsen-Olesen
magnetic vortices.

\item  The Compton wavelength phase also was discussed. We have
used the Sidharth's predictions of the non-commutativity for these
non-differentiable manifolds with aim to prove that cosmological
constants are zero, or almost zero.

\item We considered a general theory recently developed by
B.G. Sidharth and A. Das of the phase transition between the two
different lattice structures. This theory was applied to the phase
transition between the Planck scale phase and Compton scale
phase.

\item  The link between the gravitation and electromagnetism via
Dark Energy also was established by Sidharth in his recent paper.

\item We showed that for energies higher than Electroweak scale,
the analysis of the vacuum stability is reduced to the study of
the renormalization group evolution of the Higgs quartic coupling
$\lambda$. The prediction for the mass of the Higgs boson was
improved by the calculation of the two-loop radiative corrections
to the effective Higgs potential. The prediction of Higgs mass
$129.4 \pm 1.8$ GeV by Degrassi et al. provided the theoretical
explanation of the value $M_H \simeq 125.7$ GeV observed at the
LHC. Buttazzo et al. extrapolated the Standard Model parameters up
to the high (Planck scale) energies with full three-loop NNLO RGE
precision.

\item It was shown that the observed Higgs mass $M_H = 125.66 \pm 0.34$
GeV leads to a negative value of the Higgs quartic coupling $\lambda$ at
some energy scale below the Planck scale, making the Higgs
potential unstable or metastable. With the inclusion of the
three-loop RG equations, the instability scale occurs at $10^{11}$ GeV
(well below the Planck scale) meaning that at that scale the
effective potential starts to be negative, or that a new minimum
with negative cosmological constant can appear.

\item  It was shown that the experimental value of the Higgs mass
leads to a scenario which gives a borderline between the absolute
stability and metastability.

\item We assumed that the recently discovered at the LHC new
resonances with masses $m_S \simeq 750$ GeV are a new scalar $S$
bound state $6t + 6\bar t$, earlier predicted by C.D. Froggatt,
H.B. Nielsen and L.V. Laperashvili. It was shown that this bound
state, 6 top and 6 anti-top, which we identify with the 750 GeV
new boson, can provide the vacuum stability and the exact accuracy
of the Multiple Point Principle, according to which the two vacua
existing at the Electroweak and Planck scales are degenerate.

\item We calculated the main contribution of the $S$-resonance to
the effective Higgs quartic coupling $\lambda$, and showed that
the resonance with mass $m_S \simeq 750$ GeV, having the radius
$r_0 = b/m_t$ with $b \approx 2.34$, gives the positive
contribution to $\lambda$ , equal to the $\lambda = + 0.01$. This
contribution compensates the negative value of the $\lambda = -
0.01$, which was earlier obtained by Degrassi et al., and
therefore {\bf transforms the metastability of the Electroweak
vacuum into the stability.}
\end{enumerate}

\appendix

\section{Appendix: Multiple Point Principle, literature}

\section{Appendix: Non-commutativity, the main references}

\section{Appendix: The vacuum stability problem has a long history:}

\section{Appendix: Theory of the new bound state 6t + 6anti-t}

\section{Appendix: Resonance 750 GeV}
Recently both the ATLAS and CMS collaborations observed an excess
in diphoton events with an invariant mass in the region of 750
GeV. Although more data is needed to confirm or exclude the
excess, a large number of works have already appeared on this
possible new physics signal:

\newpage

\begin{figure}[H]
\centering
\includegraphics[scale=0.56]{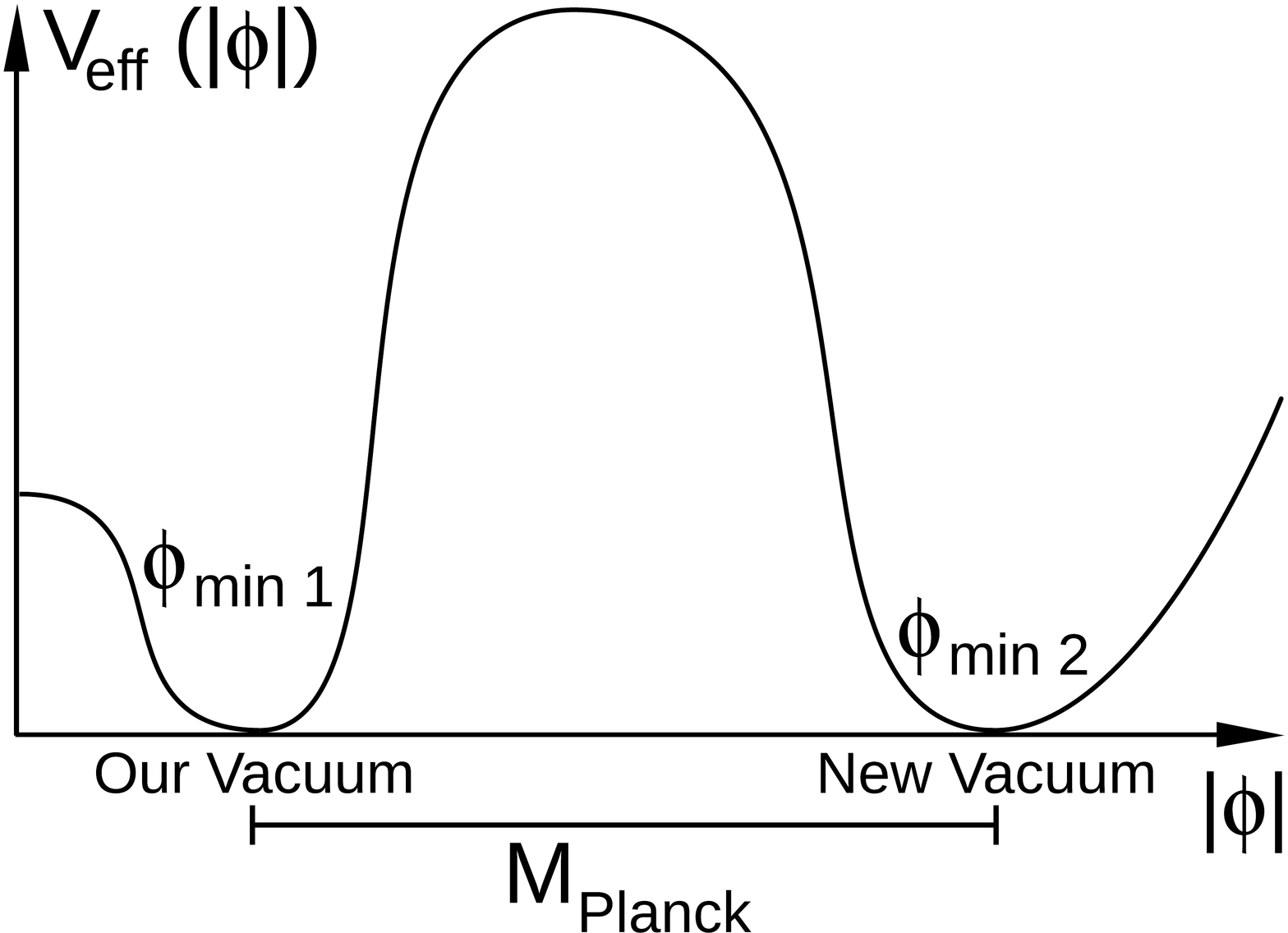}
\caption
{The second vacuum of the effective Higgs potential is degenerated with an usual Electroweak vacuum.
The Standard Model is valid up to the Planck scale except $\phi_{min2} \simeq M_{Pl}$.}
\end{figure}

\newpage

\begin{figure}[H]
\centering
\includegraphics[scale=0.5]{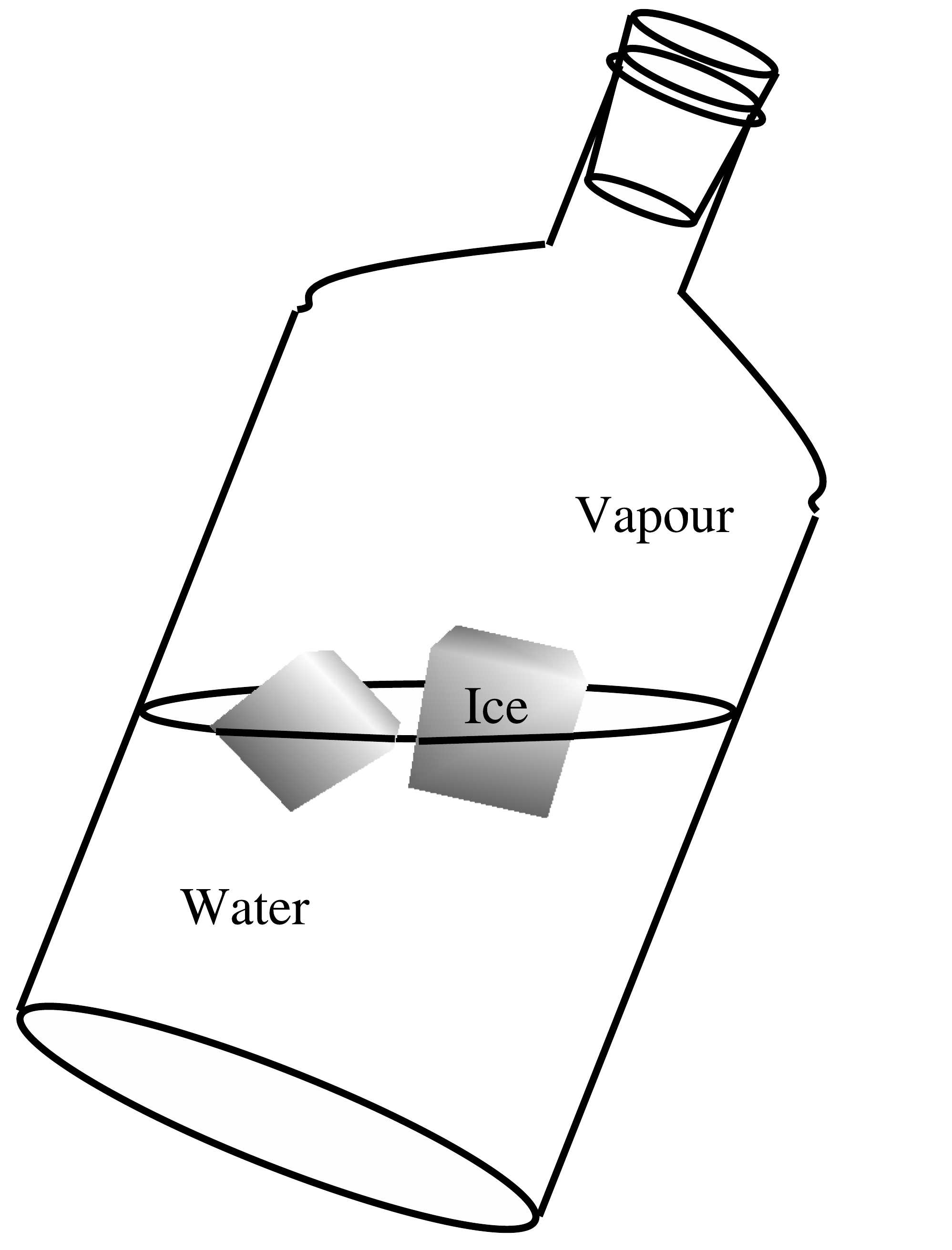}
\caption
{The degenerate phases of water (namely, ice, water and vapour) with fixed
extensive quantities: volume, energy and a number of moles.}
\end{figure}

\newpage

\begin{figure}[H]
\centering
\includegraphics[scale=1.0]{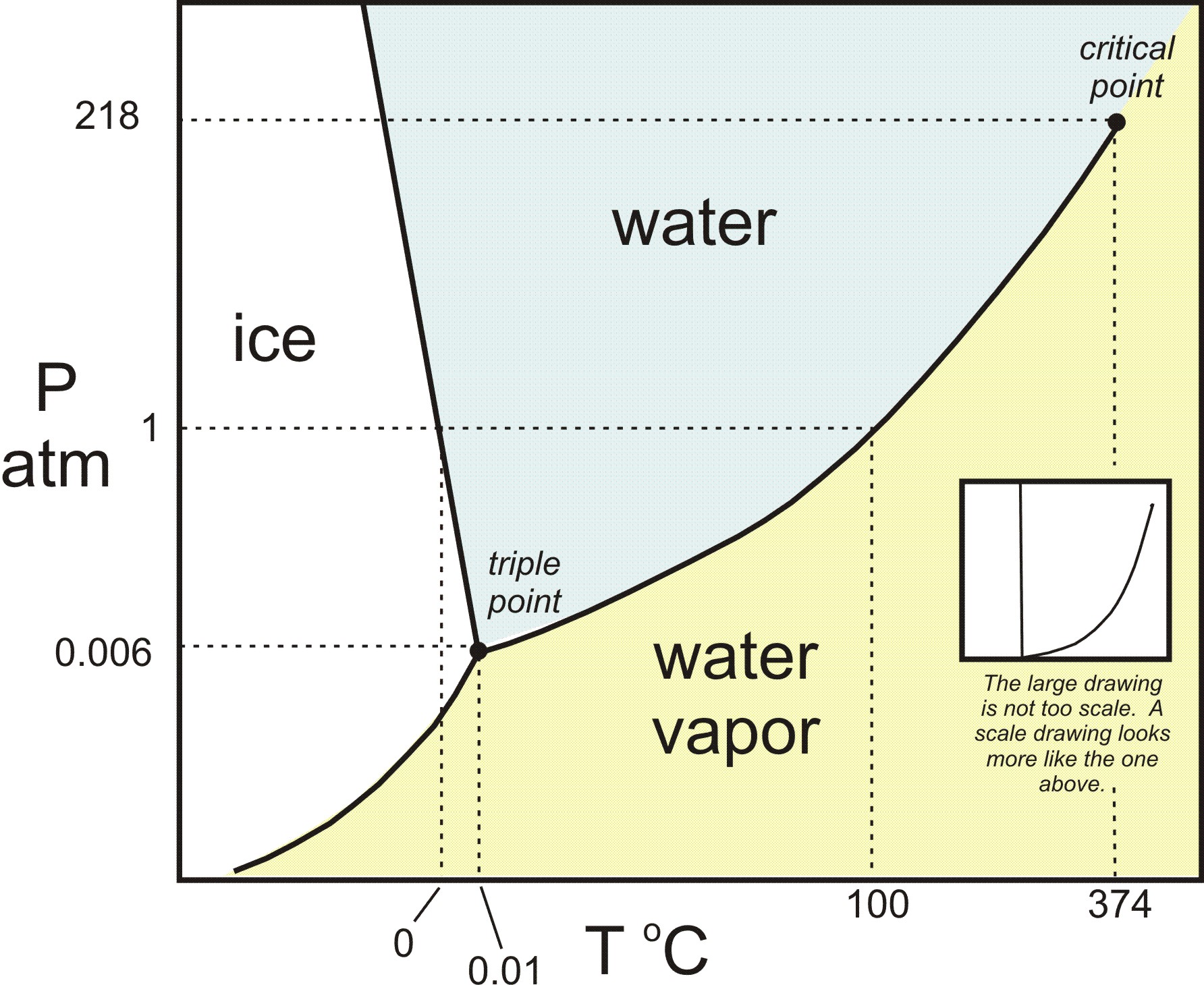}
\caption
{The phase diagram $(P,\; T)$ of water analogy. The triple point $O$ with
$T_c = 0.01 ^C$ and $P_c = 4.58\; mm\; Hg$ is shown in Fig.~2.}
\end{figure}

\newpage

\begin{figure}[H]
\centering
\includegraphics[scale=0.1235]{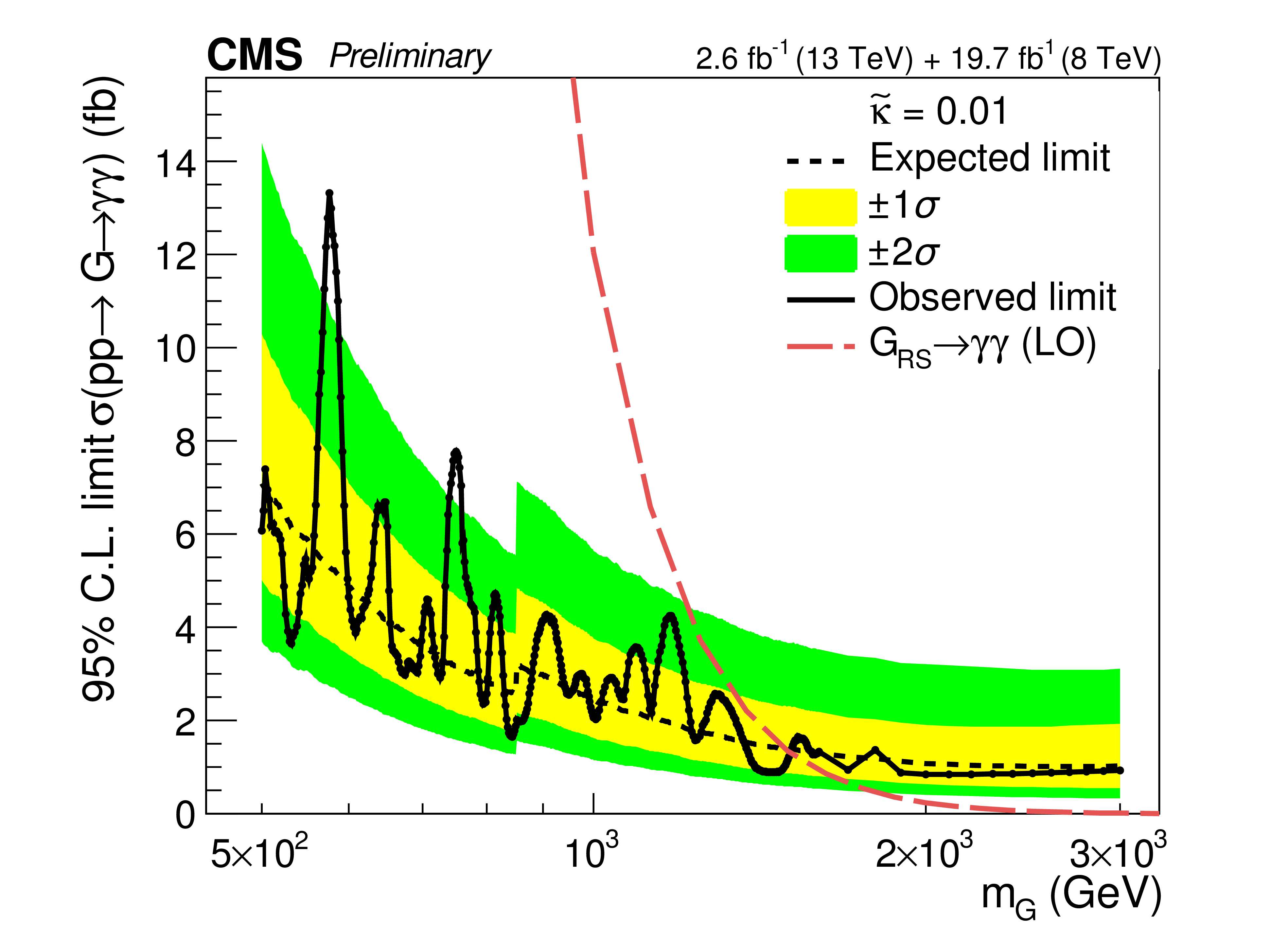}
\caption
{This figure presents searches for a new physics in
high mass diphoton events in proton-proton collisions at 13 TeV.
ATLAS and CMS Collaborations show a new resonance in the diphoton
distribution at an invariant mass of 750-760 GeV.}
\end{figure}

\newpage

\begin{figure}[H]
\centering
\includegraphics[scale=0.23]{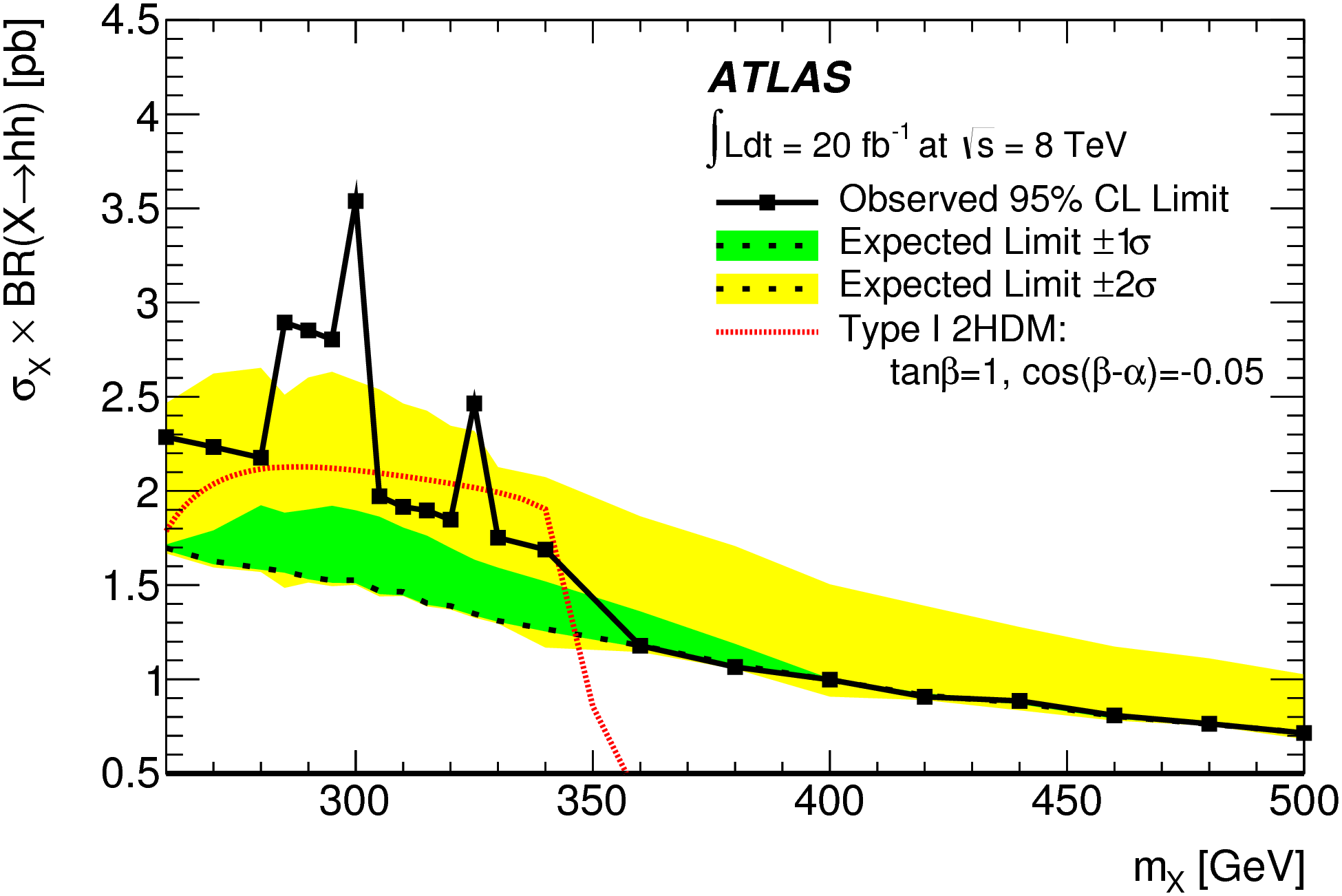}
\caption
{This figure presents searches for resonant and
non-resonant Higgs boson pair production using $20.3\; {\rm{
fb}^{-1}}$ of proton-proton collision data at $\sqrt s = 8$ TeV
generated by the LHC and recorded by the ATLAS detector in 2012.
The results show a resonance with mass $\approx 300$ GeV}
\end{figure}

\newpage

\begin{figure}[H]
\centering
\includegraphics[scale=0.66]{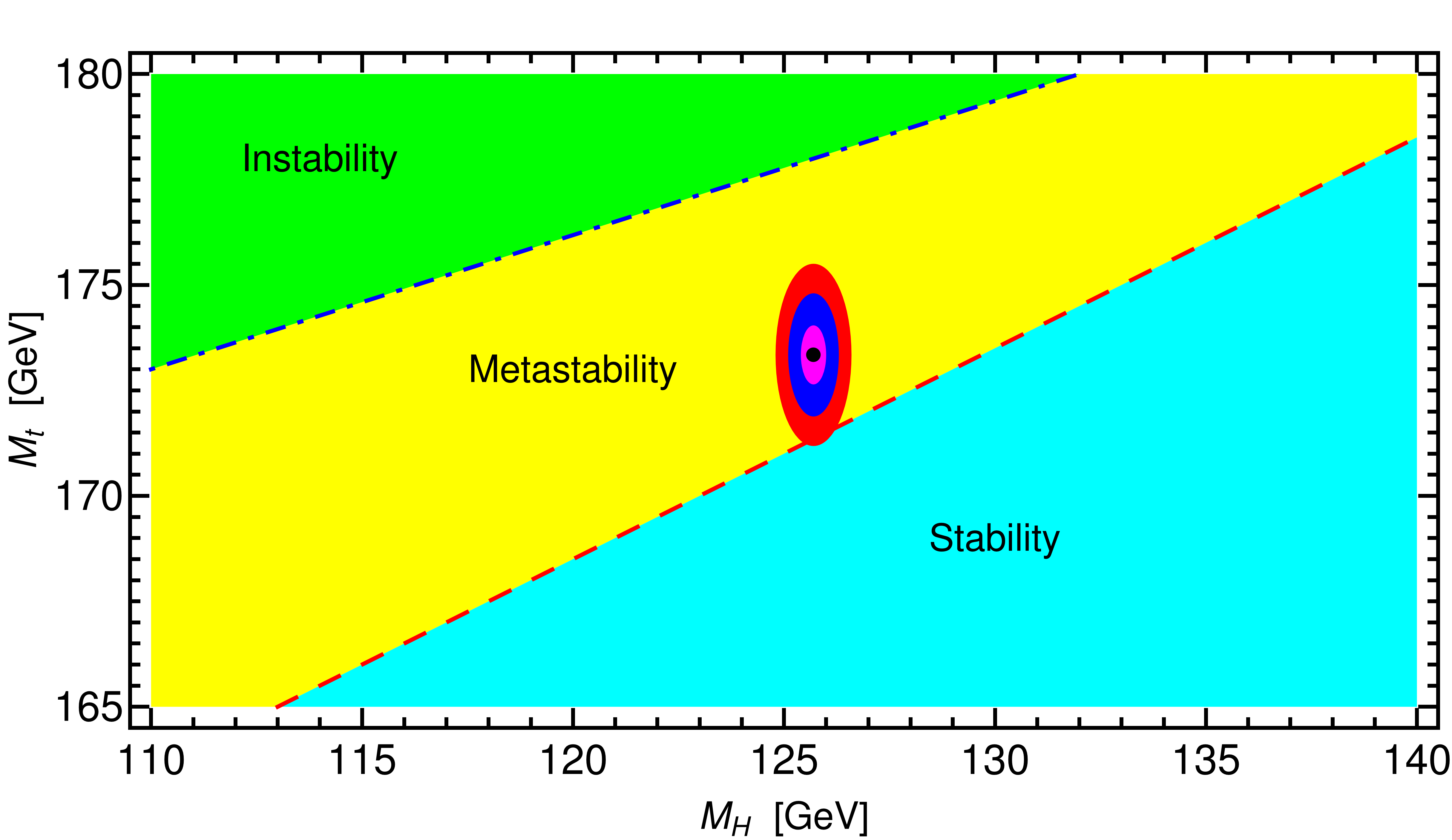}
\caption
{Stability phase diagram $(M_H,\; M_t)$ is divided into three
different sectors: 1) an absolute stability region -- cyan region of figure;
2) a metastability (yellow) region, and
3) an instability (green) region. The black dot indicates current experimental
values $M_H \simeq 125.7$ GeV and $M_t\simeq 173.34$ GeV. The ellipses take into account
$1\sigma,\; 2\sigma$ and $3\sigma$, according to the current experimental errors.}
\end{figure}

\newpage

\begin{figure}[H]
\centering
\includegraphics[scale=0.66]{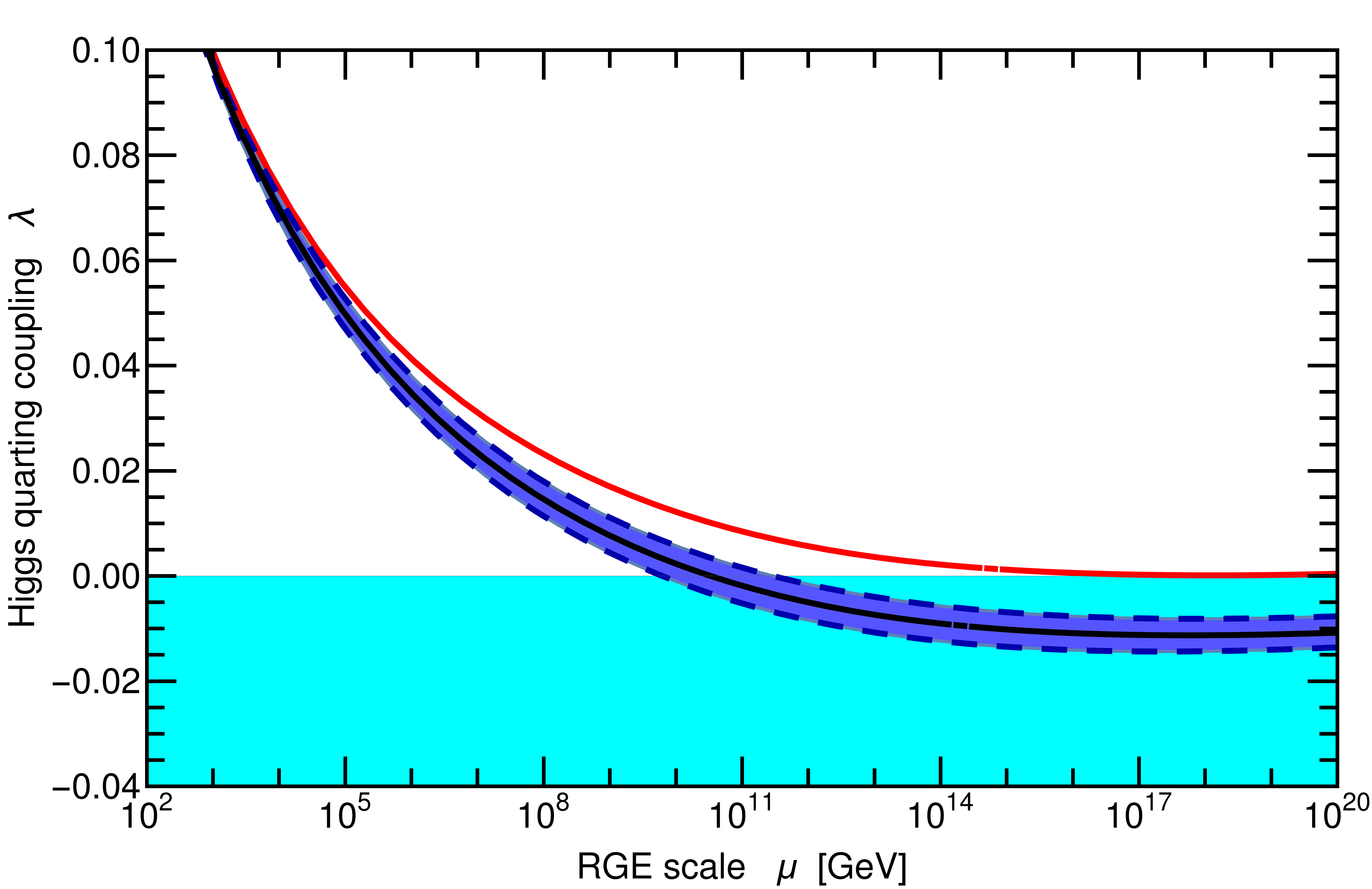}
\caption{The RG evolution of the Higgs self-coupling $\lambda(\mu)$
is given by blue lines, thick and dashed, for the current experimental values $M_H\simeq125.7$ GeV and
$M_t \simeq 173.34$ GeV for QCD constant $\alpha_s$ given by $\pm 3\sigma$.
The thick blue line corresponds to the central value of $\alpha_s = 0.1184$ and
dashed blue lines correspond to errors of  $\alpha_s$ equal to $\pm 0.0007$.}
\end{figure}

\end{document}